\documentclass[seceq]{ptptex}

\usepackage{graphicx}
\usepackage{wrapft}

\def\be{\begin{equation}}
\def\ee{\end{equation}}
\def\bea{\begin{eqnarray}}
\def\eea{\end{eqnarray}}

\title{
String Theory, Space-Time Non-Commutativity and Structure Formation%
}

\author{
Robert H. \textsc{Brandenberger}%
}

\inst{
Physics Dept., McGill University, 3600 rue Universit\'e, Montr\'eal, QC, H3A 2T8, Canada
}

\abst{
A natural consequence of string theory is a non-commutative structure of
space-time on microscopic scales. The existence of a minimal length,
and a modification of the effective field theory are two consequences of
this space-time non-commutativity. I will first explore some consequences
of the modifications of the effective field theory for structure formation in
the context of an inflationary cosmology. Then, I will explore the possibility
that the existence of a minimal length will lead to a structure formation
scenario different from inflation. Specifically, I will discuss 
recent work on string gas cosmology.
}

\begin{document}

\maketitle

\section{Introduction and Summary}

Non-commutativity of space-time on small scales appears to be one of the key
consequences of string theory. Since cosmology provides what is probably the
best way to test the consequences of new ultraviolet physics, I will in this
talk explore some possible cosmological consequences of stringy
space-time non-commutativity and of other key features of string theory.

If our universe underwent a period of cosmological inflation, then, since
the physical wavelengths of scales we observe at the present time in
observational cosmology started out smaller than the Planck length at the
beginning of the period of inflation, it should be expected that the
cosmological perturbations contain information about the ultraviolet
completion of physics \cite{RHBrev1}. It has, in fact, been shown 
\cite{Jerome} that the results of the standard theory of cosmological perturbations
(see e.g. \cite{MFB} for an in-depth review, and \cite{RHBrev2} for
a pedagogical overview) are not robust against modifications in the laws
of physics on Planck scales. In Section 3 (following a preliminary
section containing a brief summary of the essentials of the theory
of cosmological perturbations), I will summarize
work of \cite{Ho} and \cite{Tsujikawa} exploring the consequences of
space-time non-commutativity on the spectrum of cosmological fluctuations.
The main result is that, in the context of power-law accelerated expansion,
the index of the spectrum changes: a spectrum which is slightly red according
to the usual calculations becomes slightly blue on sufficiently large scales
if space-time non-commutativity is taken into account.

Conventional scalar field-driven inflationary models, however, suffer
from several conceptual problems (see e.g. \cite{RHBrev3,RHBrev4} for
recent discussions). This motivates the search for cosmological backgrounds
which do not require an inflationary period in order to explain the
observational data. In Section 4 of this article, I will discuss a
cosmological background \cite{BV} (``string gas cosmology") which emerges 
if we make use of key  stringy
degrees of freedom and key symmetries of string theory.

In Section 5 I will summarize recent work 
\cite{NBV,Ali,BNPV2} showing that, under certain
conditions on the cosmological background, thermal fluctuations of a string
gas in the very early universe can generate a scale-invariant spectrum of
cosmological perturbations, hence providing an alternative to inflation
for producing such a phenomenologically successful spectrum.
  
\section{Brief Summary of the Theory of Cosmological Perturbations}

If we neglect vector perturbations, gravitational waves, and anisotropic stress
terms, the cosmological metric including fluctuations can be written in the form 
\begin{equation} \label{longit}
ds^2 \, = \, a^2 \bigl[(1 + 2 \phi) d\eta^2
- (1 - 2 \phi)dx^2 \bigr] \, ,
\end{equation}
where $\eta$ is conformal time related to the physical time variable $t$ via
$dt = a(\eta) d\eta$, $a(\eta)$ being the scale factor of the homogeneous and
isotropic background cosmology, and the comoving spatial coordinates are denoted
by $x$. The relativistic potential $\phi$ encodes the information about
the scalar metric fluctuations and depends on space and time. To
simplify the notation, we have taken the spatial background metric to be flat.

As matter we consider a scalar field $\varphi(x, \eta)$ which can
be decomposed into a background component $\varphi_0(\eta)$ and a
fluctuation $\delta \varphi (x, \eta)$. 
The action for the matter fluctuation in Fourier space is given by
\be \label{matact1}
S \, = \, \int d\eta d^3k {1 \over 2} a^{2}(\eta) 
\bigl( \varphi^{\prime}_{-k} \varphi^{\prime}_k 
- k^2 \varphi_{-k} \varphi_k \bigr) \, ,
\ee
a prime denoting the derivative with respect to conformal time.
The fields $\phi$
and $\delta \varphi$ are related via the Einstein constraint equations.
The equations of motion for the fluctuations can be obtained by inserting
the metric and matter field decompositions into the full action and
expanding to quadratic order. The resulting action must reduce to that
of a free scalar field $v$ with time-dependent mass.
The two nontrivial tasks of the lengthy \cite{MFB} computation 
of the quadratic piece $S^{(2)}$ of the action is to find
out what combination of $\varphi$ and $\phi$ yields the variable $v$
in terms of which the action has canonical kinetic term, and what the form
of the time-dependent mass is. The result is 
\begin{equation} \label{pertact}
S^{(2)} \, = \, {1 \over 2} \int d^4x \bigl[v'^2 - v_{,i} v_{,i} + 
{{z''} \over z} v^2 \bigr] \, ,
\end{equation}
where the canonical variable $v$ (the ``Sasaki-Mukhanov variable'' introduced
in \cite{Mukh2,Mukh3,Sasaki}) is given by
\begin{equation} \label{Mukhvar}
v \, = \, a \bigl[ \delta \varphi + {{\varphi_0^{'}} \over {\cal H}} \phi
\bigr] \, 
\end{equation}
and 
\begin{equation} \label{zvar}
z \, = \, {{a \varphi_0^{'}} \over {\cal H}} \, .
\end{equation}
The variable $v$ is related to the curvature fluctuation ${\cal R}$ in comoving
coordinates via $v = z {\cal R}$.

In momentum space, the equation of motion which follows from the 
action (\ref{pertact}) is
\begin{equation} \label{pertEOM1}
v_k^{''} + k^2 v_k - {{z^{''}} \over z} v_k \, = \, 0 \, ,
\end{equation}
where $v_k$ is the k'th Fourier mode of $v$. It follows
that fluctuations oscillate on sub-Hubble scales (when $z^{''}/z$ is
negligible), but freeze out and
are squeezed $v \sim z$ on super-Hubble scales. If we choose vacuum
initial conditions at an initial time $\eta_i$, then
\begin{eqnarray} \label{incond}
v_k(\eta_i) \, = \, {1 \over {\sqrt{2 k}}} \\
v_k^{'}(\eta_i) \, = \, {{\sqrt{k}} \over {\sqrt{2}}} \, . \nonumber
\end{eqnarray}
The power spectrum of curvature perturbations, defined via
\be
P_k \, = {{k^3} \over {2 \pi^2}} {{|v_k|^2} \over {z^2}} \, ,
\ee
can then easily be calculated.
 
Note that gravitational waves and modes of a free scalar field on a fixed
cosmological background obey a similar equation of motion, but with the
function $z(\eta)$ replaced by the scale factor $a(\eta)$. If the
equation of state of matter is constant in time, then $z(\eta)$ and
$a(\eta)$ are proportional, but during a transition in the equation of state of
matter, they are no longer proportional. In the standard example of such
a transition, namely the transition during the phase of reheating in inflationary
cosmology, $z(\eta)$ changes by a much larger factor than $a(\eta)$, thus
leading to a larger squeezing for scalar cosmological fluctuations than for
gravitational waves.
  
\section{Space-Time Non-Commutativity and Inflationary Cosmology}

In this section we will summarize the results of \cite{Ho} in which
the implications of space-time non-commutativity for the spectrum of
cosmological fluctuations is explored. For consequences of space-space
non-commutativity, the reader is referred to \cite{SSnoncom}.

We start from the stringy space-time uncertainty relation \cite{SSUR}
\be
\Delta t \Delta x_p \, \geq \, l_s^2 \, , \label{ucrl}
\ee
where $t$ and $x_p$ are physical time and distance, respectively, and $l_s$
is the string length, and apply this relation to a cosmological
space-time with scale factor $a(t)$ in terms of which the metric is given by
\be
ds^2 \, = \, dt^2 - a(t)^2 dx^2 \, = 
\, a^{-2}(\tau) d\tau^2 - a(\tau)^2 dx^2 \, ,
\ee
where $\tau$ is a rescaled time.
In terms of the new time variable $\tau$, the uncertainty relation (\ref{ucrl})
can be realized by imposing the deformed commutation relation
\be
[\tau, x]_{*} \, = \, i l_s^2
\ee
where the subscript $*$ implies that the products inside the commutator are given
by the Moyal product
\be
(f * g)(x, \tau) \, = \, e^{-{i \over 2} l_s^2 (\partial_x \partial_{\tau'}
- \partial_{\tau} \partial_y)} f(x, \tau) g(y, \tau')
\ee
(evaluated at $y = x$ and $\tau' = \tau$).

The uncertainty relation has two main effects on the cosmological perturbations.
Firstly, it leads to an upper cutoff $k_c$ on the 
comoving momentum of fluctuation modes:
\be
k_c(\tau) \, = \, {{a_{eff}(\tau)} \over {l_s}} \, ,
\ee
where $a_{eff}$ is the scale factor smeared out over a time corresponding to
the space-time uncertainty:
\be
a^2_{eff}(\tau) \, = \, \bigl( {{\beta^+_k} \over {\beta^-_k}}\bigr)^{1/2}
\ee
with
\be
\beta^{\pm} \, = \, {1 \over 2} \bigl[a^{\pm 2}(\tau + kl_s^2) 
+ a^{\pm 2}(\tau - kl_s^2) \bigr] \, .
\ee

Secondly, the uncertainty relation introduces a coupling between
background geometry and fluctuations which is non-local in time, the
non-locality being a consequence of the uncertainty in time. The
theory of cosmological fluctuations in a non-commutative space-time
can be obtained \cite{Ho} from the theory in commutative space-time
by taking the action (\ref{pertact}) of cosmological perturbations
and replacing the product operators by the Moyal product. For a
real scalar field $\varphi$ on the background space-time, 
the modified action is \cite{Ho}
\be
{\tilde S} \, = \, \int d\tau dx {1 \over 2} 
\bigl( \partial_{\tau}\varphi^{\dagger} \star a^2 \star \partial_{\tau}\varphi
- \partial_x\varphi^{\dagger} \star a^{-2} \star \partial_x\varphi \bigr) \, .
\ee
which in momentum space becomes \cite{Ho}
\be \label{matact2}
{\tilde S} \, = \, \int_{k < k_c} d\eta d^3k {1 \over 2} a_k^2(\tilde \eta)
\bigl( \varphi^{\prime}_{-k} \varphi^{\prime}_k 
- k^2 \varphi_{-k} \varphi_k \bigr) \, ,
\ee
where a prime indicates the derivative with respect to ${\tilde \eta}$,
\be
a_k(\eta) \, = \, a(\eta) y_k(\eta) \,\,\,  
(y_k \, \equiv \, (\beta^-_k \beta^+_k)^{1/4}) \, ,
\ee
and $d{\tilde \eta} = a_{eff}^{-2} d\tau$. 

In the case of scalar metric
fluctuations, the ``smeared'' function $a_k$ in (\ref{matact2})
is replaced by a smeared
function $z_k$ constructed in the same way from the ``un-smeared''
function $z$ as $a_k$ is from $a$. Thus, the net effect of
non-commutativity on the evolution of cosmological fluctuations is
that the function $z$ in the equation of motion (\ref{pertEOM1})
gets replaced by the smeared function $z_k$, yielding
\be \label{pertEOM2}
v_k^{''} + k^2 v_k - {{z_k^{''}} \over {z_k}} v_k \, = \, 0 \, .
\ee
 
We will be considering a background which yields power law inflation
$a(t) \sim t^n$ with $n > 1$, where the constant $n$ is related to the
equation of state parameter $w = p / \rho$ ($p$ and $\rho$ being
pressure and energy density, respectively) via
\be
n \, = \, {2 \over {3(1 + w)}} \, .
\ee

Let us first consider UV modes, modes which are generated inside the
Hubble radius. For these modes the value of $\Delta \tau$ is small
in the sense that all smeared versions of the scale factor can be replaced by
the scale factor itself. In this case, the analysis reduces to the
usual analysis in the case of commutative space-time. The power
spectrum of scalar metric fluctuations has a red tilt
\be
P_k \, \sim \, k^{2/(1 - n)} \, .
\ee

For IR modes, on the other hand, space-time non-commutativity leads
to results which differ from the usual analysis. For these modes
the space-time uncertainty relation is saturated at a time $\tau_k^0$
when the Hubble radius is smaller than the physical wavelength.
At the time of saturation, the smearing of the scale factor is
important. Effectively, the fluctuations evolve as if they
had been generated not at the time $\tau_k^0$, but 
$\tau_k^0 + k l_s^2$. Thus, there is less squeezing of the
fluctuations then there would have been if the generation
time had been $\tau_k^0$. This effect turns the red spectrum
for UV modes into a blue spectrum for IR modes \cite{Ho}
\be
P_k \, \sim \, k^{3/(1 + n)} \, .
\ee
For fixed value of $n$, the amount of the blue tilt for the $IR$
modes is identical to the amount of the red tilt for the $UV$
modes. 

The change of the spectral tilt from red (on small angular
scales) to blue (on large angular scales) could be used to
explain the observed lack of power \cite{COBE,WMAP} in the observed angular
power spectrum of CMB fluctuations on the largest scales
\cite{Tsujikawa,Li}.

Another interesting observation which follows from the work
of \cite{Ho} is the following: if we had treated the
evolution of cosmological fluctuations with the un-modified
perturbation equation (\ref{pertEOM1}) instead of with
the modified equation (\ref{pertEOM2}), then - given vacuum
initial conditions for the modes at the time when the physical
wavelength is equal to a fixed cutoff scale - we would have
obtained a scale-invariant spectrum for the IR modes for
any expanding space-time, even if it is not accelerating.
This point was stressed subsequently in \cite{Wald}.

It is important, however, to keep in mind that the physics
which yields the minimal physical wavelength for the fluctuations
will likely also effect the initial evolution, as it does in
the non-commutative scenario discussed above. A further
worry \cite{KLM} concerning the model of \cite{Wald} is that
fluctuations on scales of observational interest to us today
are generated when the Hubble radius is much smaller than the
cutoff scale. It is hence possible that the cosmological background
used is no longer consistent at the relevant early times.

\section{Minimal Length and String Gas Cosmology}

Since string theory is based on the quantization of extended
objects, one expects string theory to give rise to the concept of
a minimal observable length whose numerical value should be related 
to the string length $l_s$. For example, the location of the
scattering point for the process of two initial strings
scattering into two final strings is smeared out over the length
scale of the string, and this is the reason that the cross-section
for string scattering does not have the UV divergence which point
particle scattering is affected with.

String gas cosmology \cite{BV} (see also \cite{Perlt} for early
work and \cite{RHBrev5,BattWat} for reviews) is a toy model of 
string cosmology in which a minimal length
scale clearly emerges. String gas cosmology is given by a background
space-time (whose action will be discussed shortly) coupled to
a gas of strings. Due to the extended nature
of strings, there are degrees of freedom which do not exist in point
particle theories. These degrees of freedom lead to a
new symmetry specific to string theory, ``T-duality", a
symmetry which implements the idea of a minimal length.

Taking all spatial directions to be toroidal
(radius $R$), strings have three types
of states: {\it momentum modes} which represent the center
of mass motion of the string, {\it oscillatory modes} which
represent the fluctuations of the strings, and {\it winding
modes} counting the number of times a string wraps the torus.
Both oscillatory and winding states are new features of string theory.

The energy of an oscillatory mode is independent of $R$, momentum
mode energies are quantized in units of $1/R$, i.e.
$E_n \, = \, n {1 \over R}$
whereas the winding mode energies are quantized in units of $R$, i.e.
$E_m \, = \, m R$
(both $n$ and $m$ are integers). Thus, we see that the energy spectrum of
the string states admits a stringy symmetry called ``T-duality 
symmetry", namely a symmetry under the change
$ R \, \rightarrow \, 1/R$
(in units of the string length $l_s$).
Under this change, the energy spectrum of string states is
not modified if $n$ and $m$ are interchanged.
The string vertex operators are consistent with this symmetry. 
Postulating
that T-duality extends to non-perturbative string theory leads
\cite{Pol} to the need of adding D-branes to the list of fundamental
objects in string theory. With this addition, T-duality is expected
to be a symmetry of non-perturbative string theory.
Specifically, T-duality will take a spectrum of stable Type IIA branes
and map it into a corresponding spectrum of stable Type IIB branes
with identical masses \cite{Boehm}.

It was argued in \cite{BV} that this symmetry leads to a minimal length.
The argument goes as follows:
Any physical detector will measure length in terms of the light
degrees of freedom. At large values of $R$, these are the momentum
modes, at small values of $R$ it is the winding modes. In terms of
winding modes, the physics on a torus of radius $1/R$ is the same
as the physics looks on a torus of radius $R$ in terms of momentum
modes. Thus, the measured length $l_{phys}$ will satisfy
$l_{phys}(1/R) \, = \, l_{phys}(R)$
and will therefore always be greater or equal to the string length.

A second important consequence of the stringy degrees of freedom for
cosmology arises from the existence of a limiting temperature for
a gas of strings in thermal equilibrium, the  {\it Hagedorn
temperature} $T_H$ \cite{Hagedorn}. Its existence follows from the
fact that the number of string oscillatory modes increases exponentially
as the string mode energy increases, there is a maximal temperature
of a gas of strings in thermal equilibrium. 
Taking a box of strings
and compressing it, the temperature will never exceed $T_H$. In fact,
as the radius $R$ decreases below the string radius, the temperature
will start to decrease, obeying the duality relation \cite{BV}
$T(R) \, = \, T(1/R)$. Figure 1 provides
a sketch of how the temperature $T$ changes as a function of $R$.
\begin{figure}
\centerline{\includegraphics[width=6 cm,height=4 cm]
                                   {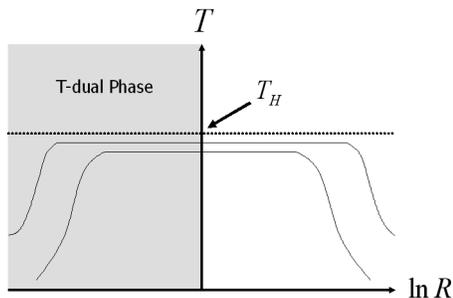}}
\caption{Sketch (based on the analysis of \cite{BV})
of the evolution of temperature $T$ as a function
of the radius $R$ of space of a gas of strings in thermal
equilibrium. The top curve is characterized by an entropy
higher than the bottom curve, and leads to a longer
region of Hagedorn behavior.}
\label{fig:2}       
\end{figure}
The conclusion from this analysis is that no temperature
singularity is expected even if the mathematical background
approaches the singular point $R \rightarrow 0$. 

To obtain the dynamics of string gas cosmology, we need to
specify an action for the background. This action must be
consistent with the basic symmetries of string theory.
Assuming, initially, that we are in the region of weak string
coupling, the action can be taken to be that of dilaton
gravity. The total action then is
\be
S \, = \, {1 \over {2 \kappa^2}} \int d^{10}x \sqrt{-{\hat g}} e^{-2 \phi}
\bigl[{\hat R} + 4 \partial^{\mu} \phi \partial_{\mu} \phi \bigr] + S_m \, ,
\ee
where ${\hat g}$ is the determinant of the metric, 
${\hat R}$ is the Ricci scalar,
$\phi$ is the dilaton, $S_m$ is the matter action (the hydrodynamical
action of a gas of strings), and
$\kappa$ is the reduced gravitational constant in ten dimensions.
The metric appearing in the above action is the metric in the
string frame. 

In the case of a homogeneous
and isotropic background given by
\be
ds^2 \, = \, dt^2 - a(t)^2 d{\bf x}^2 \, ,
\ee
the three resulting equations (the
generalization of the two Friedmann equations plus the equation
for the dilaton) in the string frame are
\cite{TV} (see also \cite{Ven})
\bea
-d {\dot \lambda}^2 + {\dot \varphi}^2 \, &=& \, e^{\varphi} E 
\label{E1} \\
{\ddot \lambda} - {\dot \varphi} {\dot \lambda} \, &=& \,
{1 \over 2} e^{\varphi} P \label{E2} \\
{\ddot \varphi} - d {\dot \lambda}^2 \, &=& \, {1 \over 2} e^{\varphi} E \, ,
\label{E3}
\eea
where $E$ and $P$ denote the total energy and pressure, respectively,
$d$ is the number of spatial dimensions, and we have introduced the
logarithm of the scale factor 
\be
\lambda(t) \, = \, {\rm log} (a(t))
\ee
and the rescaled dilaton
\be
\varphi \, = \, 2 \phi - d \lambda \, .
\ee

From the second of these equations it follows immediately
that a gas of strings containing both stable winding and
momentum modes will lead to the stabilization of the
radius of the torus: windings prevent expansion, momenta
prevent the contraction. The right hand side of the equation
can be interpreted as resulting from a confining potential for
the scale factor. From this argument it immediately follows
that a gas of strings containing both winding and momentum
modes about the compact spatial dimensions leads to a
stabilization of the radion modes (the radii of the extra
dimensions) in the string frame \cite{Watson}. Since the dilaton
is not fixed in general, this does not correspond to a fixed
radion in the Einstein frame. However, in the case of a gas of
heterotic strings, it can be shown that enhanced symmetry states
containing both momentum and winding modes lead to a 
stabilization of both volume \cite{Subodh} and shape \cite{Edna}
moduli (see also \cite{Watson2}). For a review of moduli
stabilization in string gas cosmology, the reader is referred to
\cite{RHBrev6}.

\section{String Gas Cosmology and Structure Formation}

The equations of string gas cosmology discussed in the previous
section lead to a new cosmological background. We begin in
a Hagedorn phase during which the pressure vanishes (the positive
pressure of the momentum modes canceling against the negative
pressure of the winding modes) and thus the string frame scale
factor is quasi-static. The decay of string winding modes into
ordinary matter (string loops) will lead to a smooth transition
to the radiation phase of standard cosmology. The dilaton
comes to rest in the radiation phase. For reasons discussed
in \cite{Betal} (see also \cite{KKLM}) we need to assume
that at early times in the Hagedorn phase ($t < t_R$) the dilaton
is fixed. In the later stage of the Hagedorn phase which is
described by the dilaton gravity action of the previous section
the dilaton is running (decreasing with time).

This cosmological background yields the space-time diagram 
sketched in Figure 2. For times $t < t_R$, 
we are in the static Hagedorn phase and the Hubble radius is
infinite. For $t > t_R$, the Einstein frame 
Hubble radius is expanding as in standard cosmology. The time
$t_R$ is when the dilaton starts to decrease. At a slightly
later time, the string winding modes decay, leading to the
transition to the radiation phase of standard cosmology. 
 
Given the assumptions specified above, string gas cosmology can 
lead to a causal mechanism of structure 
formation. We must compare the physical 
wavelength corresponding to a fixed comoving scale  
with the Einstein frame Hubble radius $H^{-1}(t)$. 
Recall that the Einstein frame Hubble radius separates 
scales on which fluctuations oscillate (wavelengths smaller than 
the Hubble radius) from wavelengths on which the fluctuations are frozen 
in and cannot be affected by micro-physics. Causal micro-physical 
processes can generate fluctuations only on sub-Hubble scales. 
The first 
key point is that for $t < t_i(k)$, the fluctuation mode $k$ is inside
the Hubble radius, and thus a causal generation mechanism for fluctuations
is possible. The second key point is that fluctuations evolve
for a long time during the radiation phase outside the Hubble radius.
This leads to the squeezing of fluctuations which is responsible for
the acoustic oscillations in the angular power spectrum of CMB
anisotropies (see e.g. \cite{BNPV2} for a more detailed discussion
of this point).

\begin{figure}
\centerline{\includegraphics[width=6 cm,height=6 cm]
                                   {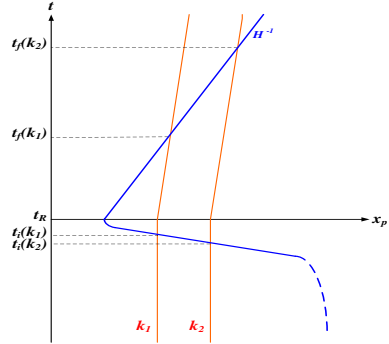}}
\caption{Space-time diagram (sketch) showing the evolution of fixed 
comoving scales in string gas cosmology. The vertical axis is time, 
the horizontal axis is physical distance.  
The solid curve represents the Einstein frame Hubble radius 
$H^{-1}$ which shrinks abruptly to a micro-physical scale at $t_R$ and then 
increases linearly in time for $t > t_R$. Fixed comoving scales (the 
dotted lines labeled by $k_1$ and $k_2$) which are currently probed 
in cosmological observations have wavelengths which are smaller than 
the Hubble radius before $t_R$. They exit the Hubble 
radius at times $t_i(k)$ just prior to $t_R$, and propagate with a 
wavelength larger than the Hubble radius until they reenter the 
Hubble radius at times $t_f(k)$.}
\end{figure}

Both in string gas cosmology and in the inflationary
scenario, fluctuations emerge on sub-Hubble scales, thus
allowing for a causal generation mechanism. However, the mechanisms
are very different. In inflationary cosmology, the exponential
expansion of space leaves behind a matter vacuum, and thus fluctuations
originate as quantum vacuum perturbations. In contrast, in string gas
cosmology space is static at early times, matter is dominated by a
string gas, and it is thus string thermodynamical fluctuations which
seed the observed structures.

The thermodynamics of a gas of strings was worked out some time
ago. We take our three spatial dimensions to be compact (specifically
$T^3$), thus admitting stable winding modes. Then, the string gas
specific heat is positive, and in this context the string thermodynamics 
was studied in detail in \cite{Deo}. 

Our procedure for string gas structure formation is the following.
For a fixed comoving scale with wavenumber $k$ we compute the matter
fluctuations while the scale in sub-Hubble (and therefore gravitational
effects are sub-dominant). When the scale exits the Hubble radius
at time $t_i(k)$ we use the gravitational constraint equations to
determine the induced metric fluctuations, which are then propagated
to late times using the usual equations of gravitational perturbation
theory (see Section 2). 

The metric including both scalar and tensor fluctuations is
\be \label{pertmetric}
d s^2 = a^2(\eta) \left\{(1 + 2 \Phi)d\eta^2 - [(1 - 
2 \Phi)\delta_{ij} + h_{ij}]d x^i d x^j\right\} \,. 
\ee 

The spectra of cosmological perturbations
$\Phi$ and gravitational waves $h$ are given by the energy-momentum 
fluctuations \cite{BNPV2}
\be \label{scalarexp} 
\langle|\Phi(k)|^2\rangle \, = \, 16 \pi^2 G^2 
k^{-4} \langle\delta T^0{}_0(k) \delta T^0{}_0(k)\rangle \, , 
\ee 
brackets indicating expectation values, and 
\be 
\label{tensorexp} \langle|h(k)|^2\rangle \, = \, 16 \pi^2 G^2 
k^{-4} \langle\delta T^i{}_j(k) \delta T^i{}_j(k)\rangle \, . 
\ee 
On the right hand side of (\ref{tensorexp}), the 
average over the correlation functions with $i \neq j$ is implied,
and $h$ indicates the amplitude of the gravitational wave $h_{ij}$. 

The root mean square energy density fluctuations in a sphere of
radius $R = k^{-1}$ are given by the specific heat
capacity $C_V$ via
\be \label{cor1}
\langle \delta\rho^2 \rangle \,  = \,  \frac{T^2}{R^6} C_V \, . 
\ee 
The result for the specific heat of a gas of closed strings
on a torus of radius $R$ is \cite{Deo}
\be \label{specheat2} 
C_V  \, \approx \, 2 \frac{R^2/\ell_s^3}{T \left(1 - T/T_H\right)}\, . 
\ee 

The power spectrum of scalar metric fluctuations is given by
\bea \label{power2} 
P_{\Phi}(k) \, 
&=& \, 8 G^2 k^{-1} <|\delta \rho(k)|^2> \, 
= \, 8 G^2 k^2 <(\delta M)^2>_R \\ 
&=& \, 8 G^2 k^{-4} <(\delta \rho)^2>_R 
= \, 8 G^2 {T \over {\ell_s^3}} {1 \over {1 - T/T_H}} 
\, , \nonumber 
\eea 
where in the first step we have used (\ref{scalarexp}) to replace the 
expectation value of $|\Phi(k)|^2$ in terms of the correlation function 
of the energy density, and in the second step we have made the 
transition to position space 

The `holographic' scaling $C_V(R) \sim R^2$ is responsible for the
overall scale-invariance of the spectrum of cosmological perturbations. 
In the above equation, for a scale $k$ 
the temperature $T$ is to be evaluated at the
time $t_i(k)$. Thus, the factor $(1 - T/T_H)$ in the 
denominator is responsible 
for giving the spectrum a slight red tilt.

As discovered in \cite{BNPV1}, the spectrum of gravitational
waves is also nearly scale invariant. In the expression
for the spectrum of gravitational waves the factor $(1 - T/T_H)$
appears in the numerator, thus leading to a slight blue tilt of
the spectrum. This is a prediction with which the cosmological
effects of string gas cosmology can be distinguished from those
of inflationary cosmology, where quite generically a slight red
tilt for gravitational waves results. The physical reason for the
blue tilt is that
large scales exit the Hubble radius earlier when the pressure
and hence also the off-diagonal spatial components of $T_{\mu \nu}$
are closer to zero.

At the present time we are still lacking a good description of the
background cosmology in the Hagedorn phase. A specific higher 
derivative gravity action in which a Hagedorn phase of string matter
can be obtained with constant dilaton is given in \cite{Tirtho}.

\section*{Acknowledgements}

I wish to thank the organizers of this conference, in particular
Satoshi Watamura, for inviting me to speak
and for their hospitality in Nishinomiya and in Kyoto. I am
grateful to Pei-Ming Ho for comments on the draft of this write-up.
The research reported
here was supported in part by an NSERC Discovery Grant, by funds from the
Canada Research Chair Program, and by a FQRNT Team Grant.

%

\end{document}